\documentclass[english,conference]{IEEEtran}
\usepackage[T1]{fontenc}
\usepackage[latin9]{inputenc}
\usepackage{amsthm}
\usepackage{amsmath}
\usepackage{amssymb}
\usepackage{graphicx}

\makeatletter
\theoremstyle{plain}
\newtheorem{thm}{\protect\theoremname}
\theoremstyle{plain}
\newtheorem{prop}[thm]{\protect\propositionname}

\ifCLASSINFOpdf
\else
\fi
\usepackage{babel}

\usepackage{babel}\providecommand{\propositionname}{Proposition}
\providecommand{\theoremname}{Theorem}

\makeatother

\usepackage{babel}
\providecommand{\propositionname}{Proposition}
\providecommand{\theoremname}{Theorem}

\begin{document}

\title{Two-Way Communication with Energy Exchange}

\author{\IEEEauthorblockN{Petar Popovski} \IEEEauthorblockA{Department
of Electronic Systems\\
 Aalborg University, Denmark\\
 Email: petarp@es.aau.dk\\
} \and \IEEEauthorblockN{Osvaldo Simeone} \IEEEauthorblockA{New
Jersey Institute of Technology\\
 Newark, USA\\
 Email: osvaldo.simeone@njit.edu} }

\maketitle




\begin{abstract}
The conventional assumption made in the design of communication systems
is that the energy used to transfer information between a sender and
a recipient cannot be reused for future communication tasks. A notable
exception to this norm is given by passive RFID systems, in which
a reader can transfer both information and energy via the transmitted
radio signal. Conceivably, any system that exchanges information
via the transfer of given physical resources (radio waves, particles,
qubits) can potentially reuse, at least part, of the received resources
for communication later on. In this paper, a two-way communication system is considered that operates
with a given initial number of physical resources, referred to as
energy units. The energy units are not replenished from outside the
system, and are assumed, for simplicity, to be constant over time.
A node can either send an {}``on'' symbol (or {}``1''), which
costs one unit of energy, or an {}``off'' signal (or {}``0''),
which does not require any energy expenditure. Upon reception of a
{}``1'' signal, the recipient node {}``harvests'' the energy contained
in the signal and stores it for future communication tasks. Inner
and outer bounds on the achievable rates are derived, and shown via
numerical results to coincide if the number of energy units is large
enough. 
\end{abstract}



\section{Introduction}

The conventional assumption made in the design of communication systems
is that the energy used to transfer information between a sender and
a recipient cannot be reused for future communication tasks. A notable
exception to this norm is given by passive RFID systems, in which
a reader can transfer both information and energy via the transmitted
radio signal. Upon reception of
the radio signal from the reader, a passive RFID tag modulates information 
by backscattering the radio energy received
from the reader (see, e.g., \cite{RFID}). Another, less conventional,
example is that of a biological system in which information is exchanged
via the transmission of particles (see, e.g., \cite{Eckford}), which
can be later reused for successive communication tasks. More in general,
any system that exchanges information via the transfer of given physical
resources (radio waves, particles, qubits) can conceivably reuse,
at least part, of the received resources for later communication tasks.

\begin{figure}
\vspace{6pt}
 \centering \includegraphics[width=3.3in]{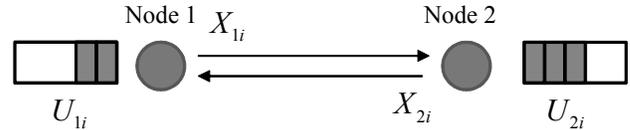} \caption{Two-way noiseless binary communication with energy exchange. The total
number of energy units is fixed (to five in the figure) and transmission
of a \textquotedbl{}1\textquotedbl{} symbol transfers energy from
the sender node to the recipient.}

\label{figsysmodel} 
\end{figure}

This paper is motivated by the examples above to consider a two-way
communication system \cite{Shannon} that operates with a given initial
number of physical resources, which we will refer to as energy units
(see Fig. \ref{figsysmodel}). The energy units are not replenished
from outside the system and can only decrease with time. To simplify
the analysis, assume that the two parties involved have a common clock
and that, at each time, a node can either send an {}``on'' symbol
(or {}``1''), which costs one unit of energy, or an {}``off''
signal (or {}``0''), which does not require any energy expenditure.
Upon reception of a {}``1'' signal, the recipient node {}``harvests''
the energy contained in the signal and stores it for future communication
tasks. In general, such harvesting process can incur an energy loss.

Furthermore, let us assume that the binary channel in either
direction is noiseless. Clearly, if there were no limitation on the
number of energy units, the nodes could communicate 1 bit per channel
use in either direction given that the channels are ideal. However,
consider now the case with a single energy unit available in
the system. Moreover, assume that there are no energy losses so that
when a {}``1'' is received, one energy unit is retrieved at the
recipient. Then, at any time instant, the energy unit is available
at either Node 1 or at Node 2. It follows that only the node that
currently possesses the energy unit can transmit a {}``1'', whereas
the other node is forced to transmit a {}``0''. Therefore, the design
of the communication strategy at the nodes should aim not only at
transferring the most information to the counterpart, but also to
facilitate energy transfer. Due to the constraints on the
available physical resources for transmission, it is expected that
the maximum sum-rate of two, achievable with no energy limitations,
cannot be attained with a limited number of energy units. It is this
trade-off that we are interested in studying in this paper.

\subsection{Contributions and Related Work}

In this paper, we will focus on the simple two-way binary noiseless
model illustrated above and assume that the initial number of energy
units is given, and can be transferred as discussed above upon transmission
of a {}``1'' symbol. We will assume, for simplicity, that there
are no energy losses in the system and investigate the set of rate
pairs achievable for any given number of energy units. Specifically,
inner and outer bounds on the achievable rates are derived, and shown
via numerical results to coincide if the number of energy units is
large enough.

A few additional remarks are in order. The model that
we consider here is different from a conventional setting in which
a conventional total average cost constraint is imposed on the two
nodes (see e.g., \cite{El Gamal Kim}). Indeed, a cost constraint
does not entail any memory in the system, whereas, in the set-up at
hand of information exchange, the current transmissions affect the
available energy in the next time instants. The model is more related
to recent works that analyze communication in the presence of energy
harvesting (see \cite{harvesting} and the references therein). However,
in those works, one assumes that the energy is harvested from the environment
in a way that is not affected by the communication process. Instead,
here, the energy available at a node depends on the previous actions
of all other nodes involved in the communication process.

\emph{Notation}: $[m,n]=\{m,m+1,...,n\}$ for integers $m\leq n$;
$\mathbb{N}$ is the set of integer numbers; Notations $H(X)$ and
$H(p(x))$ are both used to denote the entropy of a random vector
with distribution $p(x)$; If the distribution is $\verb"Bern"(p)$
we will also write $H(p)$ for the entropy. Capital letters denote
random variables and the corresponding lowercase quantities denote
specific values of the random variables.

\section{System Model}

\label{sec:SysModel}

We consider the binary and noiseless two-way system illustrated in
Fig. 1, in which the total number of \emph{energy units} in the system
is limited to a finite integer number $\verb"U"\geq1$. At any given
time instant $i$, with $i\in[1,n]$, the \emph{state} of the system
$(U_{1,i},U_{2,i})\in\mathbb{N}^{2}$ is given by the current energy
allocation between the two nodes. Specifically, a state $(U_{1,i},U_{2,i})$
indicates that at the $i$-th channel use there are $U_{j,i}$ energy
units at Node $j$, with $j=1,2$. Since we assume that $U_{1,i}+U_{2,i}=\verb"U"$
for each channel use $i\in[1,n]$ (i.e., no energy losses occur),
then we will refer to $U_{1,i}$ as the state of the system, which
always imply the equality $U_{2,i}=\verb"U"-U_{1,i}$.

At any channel use $i\in[1,n]$, each Node $j$ can transmit either
symbol $X_{j,i}=0$ or symbol $X_{j,i}=1$, and transmission of a
{}``1'' costs one energy unit, while symbol {}``0'' does not require
any energy expenditure. Therefore, the available transmission alphabet
for Node $j$, $j=1,2$ during the $i-$th channel use is 
\begin{eqnarray}
\mathcal{X}_{u}=\{0,1\} & \textrm{ if } & U_{j,i}=u\geq1 \nonumber \\
\textrm{ and }\mathcal{X}_{0}=\{0\} & \textrm{ if } & U_{j,i}=0,
\end{eqnarray}
 so that $X_{j,i}\in\mathcal{X}_{u}$ if $U_{j,i}=u$ energy units
are available at Node $j$. The channel is noiseless so that the received
signals at channel use $i$ are given by 
\begin{equation}
Y_{1,i}=X_{2,i}\textrm{ and \ensuremath{Y_{2,i}=}\ensuremath{X_{1,i}}}
\end{equation}
 for Node 1 and Node 2, respectively.

Transmission of a {}``1'' transfers one energy unit from the sender
node to the recipient node. Therefore, the state of Node 1 for $i\in[1,n]$
evolves as follows 
\begin{equation}
U_{1,i}=(U_{1,i-1}-X_{1,i-1})^{+}+X_{2,i-1},
\end{equation}
 where we set $U_{1,1}=u_{1,1}\leq\verb"U"$ as some initial state
and $U_{2,i}=\verb"U"-U_{1,i}$. We observe that the current state
$U_{1,i}$ is a deterministic function of the number $\verb"U"$ of
total energy units, of the initial state $U_{1,1}$ and of the previously
transmitted signals $X_{1}^{i-1}$ and $X_{2}^{i-1}$. We also note
that both nodes are clearly aware of the state of the system at each
time since $U_{1,i}+U_{2,i}=\verb"U"$ is satisfied for each channel
use $i$. 

Node 1 has message $M_{1}$, uniformly distributed in the set $[1,2^{nR_{1}}]$,
to communicate to Node 2, and similarly for the message $M_{2}\in[1,2^{nR_{2}}]$
to be communicated between Node 2 and Node 1. Parameters $R_{1}$
and $R_{2}$ are the transmission rates in bits per channel use (c.u.)
for Node 1 and for Node 2, respectively. We use the following definitions
for an $(n,R_{1},R_{2},\verb"U")$ code. Specifically, the code is
defined by: the overall number of energy units $\verb"U"$; two sequences
of encoding functions, namely, for Node 1, we have functions $\mathrm{f}_{1,i}$
for $i\in[1,n]$, which map the message $M_{1}$ and the past received
symbols $\ensuremath{X_{2}^{i-1}}$ (along with the initial state)
into the currently transmitted signal $X_{1,i}\in\mathcal{X}_{U_{1,i}}$;
similarly, for Node 2, we have functions $\mathrm{f}_{2,i}$ for $i\in[1,n]$,
which map the message $M_{2}$ and the past received symbols $\ensuremath{X_{1}^{i-1}}$
(along with the initial state) into the currently transmitted signal
$X_{2,i}\in\mathcal{X}_{U_{2,i}}$; and two decoding functions, namely,
for Node 1, we have a function $\mathrm{g}_{1}$, which maps all received
signals $X_{2}^{n}$ and the local message $M_{1}$ into an estimate
$\hat{M}_{1}$ of message $M_{2}$; and similarly, for Node 2, we
have a function $\mathrm{g}_{2}$, which maps all received signals
$X_{1}^{n}$ and the local message $M_{2}$ into an estimate $\hat{M}_{1}$
of message $M_{1}$. We require that the estimated messages be equal
to the true messages given the noiseless nature of the channel for
any initial state $U_{1,1}=u_{1,1}\leq\verb"U"$. 

We say that rates ($R_{1},R_{2}$) are achievable with $\verb"U"$
energy units if there exists an $(n,R_{1},R_{2},\verb"U")$ code for
all sufficiently large $n$. We are interested in studying the closure
of the set of all the rate pairs $(R_{1},R_{2})$ that are achievable
with $\verb"U"$ energy units, which we refer to as capacity region
$\mathcal{C}(\verb"U")$.

\section{Achievable Rates}

In this section, we consider various communication strategies. We
start by the simplest, but intuitively important, case with \verb"U"$=1$, and we then generalize to \verb"U"$>1$.

\subsection{\texttt{U}$=1$ Energy Unit}

We start with the special case of one energy unit ($\verb"U"=1$)
and assume the initial state $u_{1,1}=1$, so that the energy unit
is initially available at Node 1. The other case, namely $u_{1,1}=0$,
can be treated in a symmetric way. In this setting, during each channel
use, {}``information'' can be transferred only from the node where
the energy unit resides towards the other node, and not vice versa,
since the other node is forced to transmits the {}``0'' symbol.
This suggests that, when $\verb"U"=1$, the channel is necessarily
used in a time-sharing manner, and thus the sum-rate is at most one
bit per channel use. 
The first question is whether the sum-rate of 1 bit/c.u. is achievable,
and, if so, which strategy accomplishes this task.

\subsubsection{A Na\"{i}ve Strategy}

We start with a rather na\"{i}ve encoding strategy that
turns out to be insufficient to achieve the upper bound of 1 bit/c.u..
The nodes agree on a frame size $F=2^{b}>1$ channel
uses for some integer $b$ and partition the $n$ channel uses in
$n/F$ frames (assumed to be an integer for simplicity). The node
that has the energy unit at the beginning of the frame communicates
$b=\log_{2}F$ bits to the other node by placing the
energy unit in one specific channel use among the $F=2^{b}$ of the
frame. This process also transfers the energy unit to the other node,
and the procedure is repeated. The sum-rate achieved by this scheme
is 
\begin{equation}
R_{1}+R_{2}=\frac{\log_{2}F}{F}\textrm{ [bits/c.u.]},
\end{equation}
 which is rather inefficient: the maximum is achieved with
$F=2$, leading to a sum-rate of $R_{1}+R_{2}=1/2$ bits/ c.u..

The previous strategy can be easily improved by noting that the frame
can be interrupted after the channel use in which the energy unit is used, since the receiving node can still decode the transmitted $b$ bits. This strategy
corresponds to using a variable-length channel code. Specifically,
we can assign, without loss of optimality within this class of strategies,
the codeword {}``01'' to information bit {}``0'' and the codeword
{}``1'' to bit {}``1''. The average number of channel uses per
bit is thus $1/2+1/2$$\cdot2=3/2$ . Therefore, the overall number
of channel uses necessary for the transmission of $m$ bits is upper
bounded by $\frac{3m}{2}+m\epsilon$ with arbitrarily small probability
for large $m$ by the weak law of large numbers (see, e.g., \cite{El Gamal Kim}).
It follows that an achievable sum-rate is given by 
\begin{equation}
R_{1}+R_{2}=\frac{1}{3/2}=\frac{2}{3},\label{eq:RateOptimizedFrame}
\end{equation}
 which is still lower than the upper bound of 1 bit/c.u..

\subsubsection{An Optimal Strategy}

We now discuss a strategy that achieves the upper bound of $1$ bit/c.u..
The procedure is based on time-sharing, as driven by the transfer
of the energy unit from one to the other node. Specifically, each
Node $j$ has $m$ bits of information $b_{j,1},...,b_{j,m}$. Since
the initial state is $u_{1,1}=1$, Node 1 is the first to transmit:
it sends its information bits, starting with $b_{1,1}$ up until the
first bit that equals {}``1''. Specifically, assume that we have
$b_{1,1}=b_{1,2}=\cdots b_{1,i_{1}-1}=0$ and $b_{1,i_{1}}=1$. Thus,
in the $i_{1}-$th channel use the energy unit is transferred to Node
2. From the $(i_{1}+1)-$th channel use, Node 2 then starts sending
its first bit $b_{2,1}$ and the following bits until the first bit
equal to {}``1''. The process is then repeated. It is easy to see
that the total time required to finalize this two-way communication
is constant and equal to $2m$ and thus the achieved sum-rate is equal
to $R_{1}+R_{2}=1$ bit/c.u..

\subsection{$\texttt{U}>1$ Energy Units}

In the sum-capacity strategy discussed above with $\texttt{U}=1$
energy unit, both nodes transmit equiprobable symbols {}``0'' and
{}``1''. When there are $\verb"U">1$ energy units in the system,
maximizing the sum-capacity generally requires a different approach.
Consider the scenario with $\verb"U"=2$ energy units:
now it can happen that both energy units are available at
one node, say Node 1. While Node 1 would prefer to transmit equiprobable
symbols {}``0'' and {}``1'' in order to maximize the \emph{information}
flow to the recipient, one must now also consider the \emph{energy
}flow: privileging transmission of a {}``1'' over that of a {}``0''
makes it possible to transfer energy to Node 2, leading to a
state in which both nodes have energy for the next channel use. This
might be beneficial in terms of achievable sum-rate.

Based on this insight, in the following, we propose a coding strategy
that employs rate splitting and codebook multiplexing. The strategy
is a natural extension of the baseline approach discussed above for
the case $\verb"U"=1$. Each Node $j$ constructs $\verb"U"$ codebooks,
namely ${\cal C}_{j|u}$, with $u\in[1,\verb"U"]$, where codebook
${\cal C}_{j|u}$ is to be used when the Node $j$ has $u$ energy
units. Each codebook ${\cal C}_{j|u}$ is composed of codewords that
all have a specific fraction $p_{1|u}$ of {}``1'' symbols. The
main idea is that, when the number $u$ of available energy units
is large, one might prefer to use a codebook with a larger fraction
$p_{1|u}$ of {}``1'' symbols in order to facilitate energy transfer. 


\begin{prop}
\label{prop:achievable} The rate pair $(R_{1},R_{2})$ satisfying
\begin{eqnarray}
R_{1} & \leq & \sum_{u=1}^{\verb"U"}\pi_{u}H({p_{1|u})}\nonumber \\
\textrm{and }R_{2} & \leq & \sum_{u=1}^{\verb"U"}\pi_{u}H({p_{2|u})}
\end{eqnarray}
for some probabilities $0<p_{1|u},p_{2|u}<1$, $u=1\ldots\verb"U"$,
with $p_{1|0}=p_{2|\verb"U"}=0$, is included in the capacity region
$\mathcal{C}(\verb"U")$, where the probabilities $\pi_{u}\geq0,\textrm{ }u=0\ldots\verb"U"$
satisfy the fixed-point equations 
\begin{align}
\pi_{u}=\pi_{u}(\phi_{0,0|u}+\phi_{1,1|u})+\pi_{u-1}\phi_{0,1|u}+\pi_{u+1}\phi_{1,0|u}\label{eq:MarkovChainTransitions}
\end{align}
 with $\pi_{-1}=\pi_{\verb"U"+1}=0$, $\sum_{u=1}^{\verb"U"}\pi_{u}=1$,
and we have defined 
\begin{eqnarray}
\phi_{0,0|u} & = & (1-p_{1|u})(1-p_{2|\verb"U"-u})\nonumber \\
\phi_{0,1|u} & = & (1-p_{1|u})p_{2|\verb"U"-u}\nonumber \\
\phi_{1,0|u} & = & p_{1|u}(1-p_{2|\verb"U"-u})\nonumber \\
\textrm{and }\phi_{1,1|u} & = & p_{1|u}p_{2|\verb"U"-u}.\label{eq:trans_prob}
\end{eqnarray}

\end{prop}
\noindent This proposition is proved by resorting
to random coding arguments, whereby codebook ${\cal C}_{j|u}$ is
generated with independent and identically distributed (i.i.d.) $\verb"Bern"(p_{j|u})$.
As introduced above, the idea is that, when the state is $U_{1,i}=u$,
Node $j$ transmits a symbol from the codebook associated with that
state, namely codebook ${\cal C}_{1|u}$ for Node $1$ and codebook
$\mathcal{C}_{2|\verb"U"-u}$ for Node 2 (which has $\verb"U"-u$
energy units). Both nodes know the current state $U_{1,i}$
and thus can demultiplex the codebooks at the receiver side. According
to the random coding argument, the state $U_{1,i}$ evolves according
to a Markov chain: the system stays in the same state $u$ with probability
$\phi_{0,0|u}+\phi_{1,1|u}$ (both nodes transmit {}``0'' or {}``1''),
changes to the state $u+1$ with probability $\phi_{1,0|u}$ (Node
1 transmits a {}``1'' and Node 2 a {}``0'') or changes to the
state $u-1$ with probability $\phi_{0,1|u}$ (Node 1 transmits a
{}``0'' and Node 2 a {}``1''). The definition of
the conditional probabilities (\ref{eq:trans_prob}) reflects the
fact that the codebooks are generated independently by the two nodes.
A full proof is given in Appendix~\ref{sec:ProofProp1}.

\section{Outer Bounds}

\label{sec:Converse}

In this section, we derive an outer bound to the capacity region $\mathcal{C}(\verb"U")$.
To set up the notation, for a given joint distribution $\phi_{x_{1},x_{2}|u}\geq0$,
conditional on some value $u$, with $x_{1},x_{2}\in\{0,1\}$ and
$\sum_{x_{1},x_{2}\in\{0,1\}}\phi_{x_{1},x_{2}|u}=1$, we define the
marginal distributions $\phi_{x_{1}|u}=\sum_{x_{2}=0}^{1}\phi_{x_{1},x_{2}|u}$
and $\phi_{x_{2}|u}=\sum_{x_{1}=0}^{1}\phi_{x_{1},x_{2}|u}$, and
the conditional distributions $\phi_{x_{1}|x_{2},u}=\phi_{x_{1},x_{2}|u}/\phi_{x_{2}|u}$
and $\phi_{x_{2}|x_{1},u}=\phi_{x_{1},x_{2}|u}/\phi_{x_{1}|u}$, for
$x_{1},x_{2}\in\{0,1\}$.
\begin{prop}
If the rate pair ($R_{1},R_{2}$) is included in the capacity region
$\mathcal{C}(\verb"U")$, then there exist probabilities $\pi_{u}\geq0$
with $\sum_{u=1}^{\verb"U"}\pi_{u}=1$, and \textup{$\phi_{x_{1},x_{2}|u}\geq0$
with }$\sum_{x_{1},x_{2}\in\{0,1\}}\phi_{x_{1},x_{2}|u}=1$ for all
$u\in\{0,1,...,\verb"U"\}$, such that $\phi_{1,x_{2}|0}=0$ for $x_{2}\in\{0,1\}$,
$\phi_{x_{1},1|U}=0$ for $x_{1}\in\{0,1\}$, condition (\ref{eq:MarkovChainTransitions})
is satisfied, and the following inequalities hold
\begin{align}
R_{1} & \leq\sum_{u=0}^{U}\pi_{u}\sum_{x_{2}=0}^{1}\phi_{x_{2}|u}H\left(\phi_{1|x_{2},u}\right)\\
R_{2} & \leq\sum_{u=0}^{U}\pi_{u}\sum_{x_{1}=0}^{1}\phi_{x_{1}|u}H\left(\phi_{1|x_{1},u}\right)\\
\textrm{and }R_{1}+R_{2} & \leq\sum_{u=0}^{U}\pi_{u}H\left(\phi_{x_{1},x_{2}|u}\right).\label{eq:sum_rate_upper}
\end{align}
\end{prop}
\noindent The outer bound above can be interpreted as follows. Suppose
that, when the state is $U_{1,i}=u$, the nodes were allowed to choose
their transmitted symbols according to a \emph{joint} distribution
$\phi_{x_{1},x_{2}|u}=\Pr[X_{1,i}=x_{1},\textrm{ }X_{2,i}=x_{2}]$.
Note that this is unlike the achievable strategy described in the
previous section in which the codebook were generated independently.
Intuitively, allowing for correlated codebooks, leads to a larger
achievable rate region, as formalized by Proposition 2, whose proof
can be found in Appendix B.

\section{Numerical Results}

Fig.~\ref{fig:num_results} compares the achievable sum-rate obtained
from Proposition 1 and the upper bound (\ref{eq:sum_rate_upper})
on the sum-rate obtained from Proposition 2 versus the total number
of energy units $\verb"U"$. As for the achievable sum-rate, we consider
both a conventional codebook design in which the same probability
$p_{j|u}=0.5$ is used irrespective of the state $U_{1.i}=u$, and
one in which the probabilities $p_{j|u}$ are optimized. It can be
seen that using conventional codebooks, which only aim at maximizing
information flow on a single link, leads to substantial performance
loss. Instead, the proposed strategy with optimized probabilities
$p_{j|u}^{*}$, which account also for the need to manage the energy
flow in the two-way communication system, performs close to the upper
bound. The latter is indeed achieved when $\verb"U"$ is large enough.

A remark on the optimal probabilities $p_{j|u}^{*}$ is in order.
Due to symmetry, it can be seen that we have $p_{1|u}^{*}=p_{2|\verb"U"-u}^{*}$.
Moreover, numerical results show that $p_{1|u}^{*}$ increases monotonically
as $u$ goes from $0$ to \verb"U", such that $p_{1,\verb"U"}^{*}>0.5$.
In particular, when the number of states $\verb"U"+1$ is odd, it
holds that $p_{1,\verb"U"/2}^{*}=p_{2,\verb"U"/2}^{*}=0.5$. It is finally
noted that the energy neutral transitions (both nodes emitting {}``0''
or both emitting {}``1'') occur with equal probability (i.e., $(1-p_{1,u}^{*})(1-p_{2,u}^{*})=p_{1,u}^{*}p_{2,u}^{*}$). 

\begin{figure}[t]

 \centering 
 \includegraphics[width=9cm]{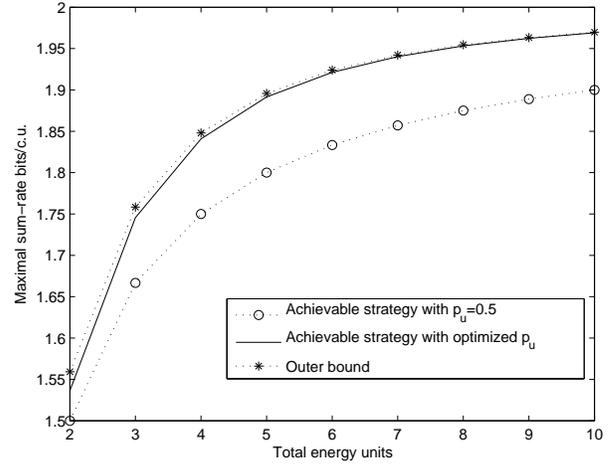} 
\caption{Achievable sum-rate obtained from Proposition 1 and upper bound 
(\ref{eq:sum_rate_upper})
versus the total number of energy units \texttt{U}. }
\label{fig:num_results}
\end{figure}

\section{Conclusion}

In resource-constrained systems in which the resources (e.g., energy)
used for communication can be {}``recycled'', new performance trade-offs
arise due the need to balance the maximization of the information
flow with the resulting resource exchange among the communicating
nodes. This paper has illustrated this aspect by studying a simple
two-way communication scenario with noiseless channels and limited
resources. Various extensions of this work are of interest, including
a generalization to a scenario with energy losses.

\appendices{}

\section{Proof of Proposition 1}

\label{sec:ProofProp1}

\subsubsection{Code construction}

We generate $\verb"U"$ codebooks for each Node $j=1,2$, namely ${\cal C}_{j|u}$,
with $u\in[1,\verb"U"]$. The codebook ${\cal C}_{j|u}$ for $u>0$
has $K_{j,u}$ codewords, each consisting of $n_{j,u}$ symbols $\tilde{x}_{j,u,l}\in\{0,1\}$,
which are randomly and independently generated as $\verb"Bern"(p_{j|u})$
variables, with $l=1,2,...,n_{j,u}$ and $n_{j,u}=n\delta_{j,u}$,
for some $0\leq\delta_{j,u}<1$. We denote the codewords as $\tilde{x}_{j,u}^{n_{j,u}}(m_{j,u})$
with $m_{j,u}\in[1,K_{j,u}]$. Note that the parameter $\delta_{j,u}$
does not depend on $n$, and hence, if $n\rightarrow\infty$, then
we have $n_{j,u}\rightarrow\infty$ for all $j,u$. We set $2^{nR_{j}}=\prod_{u=1}^{\verb"U"}K_{j,u}$,
while the relations among the remaining parameters ($K_{j,u}$,$\delta_{j,u}$,$p_{j|u}$)
will be specified below.


\subsubsection{Encoding}

Each node performs rate splitting. Namely, given a message $M_{j}\in[1,2^{nR_{j}}]$,
Node $j$ finds a $\verb"U"-$tuple $[m_{j,1},...,m_{j,\verb"U"}]$
with $m_{j,u}\in[1,K_{j,u}]$ that uniquely represents $M_{j}$. This
is always possible since we have $2^{nR_{j}}=\prod_{u=1}^{\verb"U"}K_{j,u}$.
Then, the selected codewords $\tilde{x}_{j,u}^{n_{j,u}}(m_{j,u})$
for $u\in[1,\verb"U"]$ are transmitted via multiplexing based on
the current available energy. Specifically, each Node $j$ initializes
$\verb"U"$ pointers $l_{j,1}=l_{j,2}=\cdots=l_{j,\verb"U"}=1$ that
keep track of the number of symbols already sent from codewords $\tilde{x}_{j,1}^{n_{j,1}}(m_{j,1})$,
$\tilde{x}_{j,2}^{n_{j,2}}(m_{j,2})$, ..., $\tilde{x}_{j,\verb"U"}^{n_{j,\verb"U"}}(m_{j,\verb"U"})$,
respectively. At channel use $i$, if the state is $U_{1,i}=u$, then
the nodes operate as follows. 
\begin{itemize}
\item Node 1: If $u=0$, then $x_{1,i}=0$. Else, if $l_{1,u}\leq n_{1,u}$,
Node $1$ transmits $x_{1,i}=\tilde{x}_{1,u,l_{1,u}}(m_{1,u})$ and
increments the pointer $l_{1,u}$ by 1. Finally, if $l_{1,u}=n_{1,u}+1$
the pointer $v_{1,u}$ is not incremented, and the transmitter uses
\emph{random padding}, i.e., it sends $x_{1,i}=1$ with probability
$p_{1,u}$ and $x_{1,i}=0$ otherwise. 
\item Node 2: If $u=\verb"U"$ (i.e., no energy is available at Node 2),
then $x_{2,i}=0$. Else, if $l_{2,\verb"U"-u}\leq n_{2,\verb"U"-u}$,
Node $2$ transmits $x_{2,i}=\tilde{x}_{2,\verb"U"-u,2,l_{2,\verb"U"-u}}(m_{2,\verb"U"-u})$
and increments the pointer $l_{2,\verb"U"-u}$ by 1. Finally, if $l_{2,\verb"U"-u}=n_{2,\verb"U"-u}+1$,
the pointer $l_{2,\verb"U"-u}$ is not incremented, and Node 2 sends
$x_{2,i}=1$ with probability $p_{2,\verb"U"-u}$ and $x_{2,i}=0$
otherwise. 
\end{itemize}
The random padding method used above is done for technical reasons
that will be clarified below.

\subsubsection{Decoding}

We first describe the decoding strategy for Node 2. By construction,
the nodes are aware of the state sequence $U_{1}^{n}$, and thus can
determine the ordered set 
\begin{equation}
{\cal N}_{u}=\{i|U_{1,i}=u\},
\end{equation}
 of channel use indices in which the state is $u$ with $u\in[0,$$\verb"U"${]}.
For all $u\in[1,$$\verb"U"${]}, if $|{\cal N}_{u}|\geq n_{1,u}$,
then Node 2 takes the first $n_{1,u}$ indices $i_{u,1}<i_{u,2}<\cdots<i_{u,n_{1,u}}$
from the set ${\cal N}_{u}$ and obtains the list of messages $m_{1,u}\in[1,K_{1,u}]$
that satisfy $\tilde{x}_{1,u,k}(m_{1,u})=x_{1,i_{u,k}}$ for all $k\in[1,n_{1,u}]$.
Note that the list cannot be empty due to the fact that the channel
is noiseless. However, it contains more than one message, or if $|\mathcal{N}_{u}|<n_{1,u}$,
then Node 2 puts out the estimate $\hat{m}_{1,u}=1$. Instead, if
the list contains only one message $m_{1,u}$, then Node 2 sets $\hat{m}_{1,u}=m_{1,u}$.
Finally, the message estimate is obtained as $\hat{m}_{1}=[\hat{m}_{1,1},...,\hat{m}_{1,\verb"U"}]$.

Node 1 operates in the same way, with the only caveat that the $u$th
codebook $\mathcal{C}_{2|u}$ of Node 2 is observed at channel uses
in the set ${\cal N}_{U-u}$ for $u\in[1,\verb"U"]$.

\subsubsection{Analysis}

We evaluate the probability of error on average over the messages
and the generation of the codebooks, following the random coding principle.
From the definition of the decoders given above, the event that any
of the decoders is in error is included in the set ${\cal E}=\bigcup_{j=1,2}\bigcup_{u=1}^{\verb"U"}({\cal E}_{j,u}^{(1)}\cup{\cal E}_{j,u}^{(2)})$,
where: (\emph{i}) ${\cal E}_{j,u}^{(1)}$ is the event that $|\mathcal{N}_{u}|<n_{1,u}$
for $j=1$ and that $|\mathcal{N}_{\verb"U"-u}|<n_{2,u}$ for $j=2$,
that is, that the number of channel uses in which the system resides
in the state in which the codeword $\tilde{x}_{j,u}^{n_{j,u}}(m_{j,u})$
from the codebook $\mathcal{C}_{j,u}$ is sent is not sufficient to
transmit the codeword in full; (\emph{ii}) ${\cal E}_{j,u}^{(2)}$
is the event that two different messages $m_{j,u}^{\prime},m_{j,u}^{\prime\prime}\in[1,K_{j,u}]$
are represented by the same codewords, i.e., $\tilde{x}_{j,u}^{n_{1,u}}(m_{j,u}^{\prime})=\tilde{x}_{1,u}^{n_{,1u}}(m_{1,u}^{\prime\prime})$.

The probability of error can thus be upper bounded as 
\begin{equation}
\Pr[\mathcal{E}]\leq\sum_{j=1}^{2}\sum_{u=1}^{\verb"U"}\left(\Pr[{\cal E}_{j,u}^{(1)}]+\Pr[{\cal E}_{j,u}^{(2)}]\right).\label{eq:Pje}
\end{equation}
 In the following, we evaluate upper bounds on this terms.

It immediately follows from the packing lemma of \cite{El Gamal Kim}
that $\Pr[{\cal E}_{j,u}^{(2)}]\rightarrow0$ as $n_{j,u}\rightarrow\infty$
as long as 
\begin{equation}
\frac{\log_{2}K_{j,u}}{n_{j,u}}<H(p_{j|u})-\delta(\epsilon)
\end{equation}
with $\delta(\epsilon)\rightarrow0$ as $\epsilon\rightarrow0$. For
analysis of the probabilities $\Pr[{\cal E}_{j,u}^{(1)}]$, we observe
that, under the probability measure induced by the described random
codes, the evolution of the state $U_{1,i}$ across the channel uses
$i\in[1,n]$ is a Markov chain with $\verb"U"+1$ states. Specifically,
the chain is a birth-death process, since, if the state is $U_{1,i}=u$
in channel use $i$, the next state $U_{1,i+1}$ can only be either
$u-1$ or $u+1$. More precisely, let $q_{u|w}=\Pr(U_{1,i+1}=u|U_{1,i}=w$)
be the transition probability. Note that, due to the use of random
padding, the transition probability $q_{u|w}$ remains constant during
all $n$ channel uses, so that the Markov chain is time-invariant.

We now elaborate on the Markov chain $U_{1,i}$. To this end, we first
define as $\phi_{x_{1},x_{2}|u}$, where $x_{1},x_{2}\in\{0,1\}$
be the joint probability that Node 1 transmits $X_{1,i}=x_{1}$ and
Node 2 transmits $X_{2,i}=x_{2}$ during the $i-$th channel use in
which the state is $U_{1,i}=u$. Specifically, from the way in which
the codebooks are generated, we have (). We can now write the non-zero
values of the transition probability $q_{u|w}$ as follows: 
\begin{align}
q_{u,u-1}=\phi_{1,0|u}\quad q_{u,u+1}=\phi_{0,1|u}\nonumber \\
q_{u,u}=1-q_{u,u-1}-q_{u,u+1}
\end{align}
 With a slight abuse of the notation and noting that $\phi_{1,0|0}=\phi_{1,1|0}=0$
and $\phi_{0,1|\verb"U"}=\phi_{1,1|\verb"U"}=0$ the expressions above
also represent the transitions for the two extremal states $u=0$
and $u=\verb"U"$, as they imply $q_{0|-1}=0$ and $q_{\verb"U"|\verb"U"+1}=0$.

If $p_{1,0}=p_{2,0}=0$ and $0<p_{1,u},p_{2,u}<1$ for all $u>0$,
then it can be seen that the Markov chain is aperiodic and irreducible,
and thus there exist a unique set of stationary probabilities $\pi_{0},\pi_{1},\cdots,\pi_{\verb"U"}$,
which are given by solving the linear system, defined by taking \verb"U"
equations of type (\ref{eq:MarkovChainTransitions}) for $u=0\ldots\verb"U"-1$
and adding the condition $\sum_{u=0}^{\verb"U"}\pi_{u}=1$.

We are now interested in the statistical properties of the set $|\mathcal{N}_{u}|$
of channel uses in which the state satisfies $U_{1}=u$. Using the
ergodic theorem and the strong law of large numbers \cite[Theorem 1.10.2]{Norris},
it can be shown that $\lim_{n\rightarrow\infty}\frac{V_{u}(n)}{n}=\pi_{u}$
with probability 1. Therefore, if we choose: 
\begin{equation}
l_{1,u}=l_{2,\verb"U"-u}=n(\pi_{u}-\epsilon)
\end{equation}
then $\Pr[{\cal E}_{1,u}^{(2)}]=\Pr[{\cal E}_{2,\verb"U"-u}^{(2)}]$
can be made arbitrarily close to $0$ as $n\rightarrow\infty$. This
concludes the proof.

\section{Proof of Proposition 2}

Consider any $(n,R_{1},R_{2},\verb"U")$ code with zero probability
of error, as per our definition of achievability in Sec. \ref{sec:SysModel}.
We have the following inequalities: 
\begin{align}
nR_{1} & =H(M_{1})=H(M_{1}|M_{2},U_{1,1}=u_{1,1})\nonumber \\
 & \overset{\textrm{\ensuremath{(a)}}}{=}H(M_{1},X_{1}^{n},U_{1}^{n}|M_{2},U_{1,1}=u_{1,1})\nonumber \\
 & \overset{(b)}{=}H(X_{1}^{n},U_{1}^{n}|M_{2},U_{1,1}=u_{1,1})\\
 & \overset{}{=}\sum_{i=1}^{n}H(X_{1,i},U_{1,i}|X_{1}^{i-1},U_{1}^{i-1},M_{2},U_{1,1}=u_{1,1})\nonumber \\
 & =\sum_{i=1}^{n}H(U_{1,i}|X_{1}^{i-1},U_{1}^{i-1},M_{2},U_{1,1}=u_{1,1})\\
 & +H(X_{1,i}|X_{1}^{i-1},U_{1}^{i},M_{2},U_{1,1}=u_{1,1})\\
 & \overset{(c)}{=}\sum_{i=1}^{n}H(X_{1,i}|X_{1}^{i-1},U_{1}^{i},M_{2},U_{1,1}=u_{1,1})\\
 & \overset{(d)}{\leq}\sum_{i=1}^{n}H(X_{1,i}|U_{1,i},X_{2,i})\\
 & \overset{(e)}{=}H(X_{1}|U_{1},X_{2},Q)\\
 & \leq H(X_{1}|U_{1},X_{2}),
\end{align}
where (\emph{a}) follows since $X_{1}^{n},U_{1}^{n}$ are functions
of $M_{1},M_{2}$ and $u_{1,1}$; (\emph{b}) follows since $H(M_{1}|X_{1}^{n},U_{1}^{n},M_{2},U_{1,1}=u_{1,1})=0$
holds due to the constraint of zero probability of error; (\emph{c})
follows since $U_{1,i}$ is a function of $X_{1}^{i-1},M_{2}$ and
$u_{1,1}$; (\emph{d}) follows by conditioning reduces entropy; (e)
follows by defining a variable $Q$ uniformly distributed in the set
$[1,n]$ and independent of all other variables, along with $X_{1}=X_{1Q}$,
$X_{2}=X_{2Q}$ and $U_{1}=U_{1Q}$.

Similar for $nR_{2}$ we obtain the bound $nR_{1}\leq H(X_{1}|U_{1},X_{2})$.
Moreover, for the sum-rate, similar steps lead to 
\begin{align}
n(R_{1}+R_{2}) & =H(M_{1},M_{2})=H(M_{1},M_{2}|U_{1,1}=u_{1,1})\nonumber \\
 & =H(M_{1}M_{2},X_{1}^{n},X_{2}^{n},U_{1}^{n}|U_{1,1}=u_{1,1})\nonumber \\
 & =H(X_{1}^{n},X_{2}^{n},U_{1}^{n}|U_{1,1}=u_{1,1})\nonumber \\
 & =\sum_{i=1}^{n}H(U_{1,i}|X_{1}^{i-1},X_{2}^{i-1},U_{1}^{i-1},M_{2},U_{1,1}=u_{1,1})\nonumber \\
 & +H(X_{1,i},X_{2,i}|X_{1}^{i-1},X_{2}^{i-1},U_{1}^{i},M_{2},U_{1,1}=u_{1,1})\nonumber \\
 & \geq\sum_{i=1}^{n}H(X_{1,i},X_{2,i}|U_{1,i})\nonumber \\
 & =H(X_{1},X_{2}|U_{1}).
\end{align}

Let us now define $\pi_{u}=\Pr[U_{1}=u]$ and $\phi_{x_{1},x_{2}|u}=\Pr[X_{1}=x_{1},X_{2}=x_{2}|U_{1}=u]$
for $i,j\in\{0,1\}$ and for all $u_{1}\in\{0,1,...,\verb"U"\}$.
Probability conservation implies that the relationship (\ref{eq:MarkovChainTransitions})
be satisfied. This concludes the proof.

\end{document}